\documentclass[letterpaper,twocolumn,10pt]{article}
\usepackage{usenix2019_v3}

\usepackage{tikz}
\usepackage{amsmath}

\usepackage{filecontents}

\usepackage{subcaption}
\usepackage{booktabs} 
\usepackage{xcolor}
\usepackage{enumitem}
\usepackage{xspace}
\usepackage{graphicx}

\usepackage{amssymb}    
\usepackage{booktabs}   
\usepackage{pifont}

\usepackage{xcolor}

\definecolor{okgreen}{HTML}{009E73}
\definecolor{badred}{HTML}{D55E00}
\definecolor{fbamber}{HTML}{B26A00}
\begin{document}

\date{}

\title{\Large \bf Memento: Communicating Sensor Observations Through Shared Memory}

\author{
{\rm Chae Young Lee}\\
Stanford University
\and
{\rm Lan Mei}\\
Stanford University
\and
{\rm Yifan Yang}\\
Stanford University
\and{\rm Sara Achour}\\
Stanford University
\and
{\rm Zerina Kapetanovic}\\
Stanford University
} 

  
\newcommand{\prh}[1]{\noindent\textbf{#1}}
\newcommand{\todo}[1]{\textcolor{red}{\textbf{TODO:} #1}}

\newcommand{\luke}[1]{\textcolor{blue}{\textbf{Luke:} #1}}
\newcommand{\lan}[1]{\textcolor{blue}{\textbf{Lan:} #1}}

\newcommand{\method}[1]{\textsf{#1}}
\newcommand{\tool}{Memento\xspace}
\newcommand{\short}{Memento\xspace}

\newcommand{\hw}[1]{\textcolor{blue}{#1}}

\maketitle

\begin{abstract}
Modern wireless sensing systems increasingly collect high-dimensional data such as images and video, but uplink communication remains a major bottleneck for low-power sensor deployments. Existing approaches reduce communication through compression or task-oriented features, yet they typically encode each observation independently and often rely on fixed models that require retraining under distribution shift. However, many sensing workloads exhibit strong redundancy, as devices repeatedly observe similar scenes and objects over time. We present \tool, a communication system that exploits this redundancy by transmitting observations \emph{by reference}. Instead of sending every sample in full, the sender embeds each observation and searches a shared table of previously observed samples stored in persistent memory. When a sufficiently similar match is found, the sender transmits only a compact reference to that entry. We implement \short on a prototype comprising a microcontroller, an SD card, and a Long Range radio and evaluate it across three image datasets and one radar dataset. Results show that Recurio shifts the accuracy-communication and accuracy-energy Pareto frontier outward, matching cloud accuracy with up to $26\times$ lower latency and $27\times$ lower energy at comparable accuracy.
\end{abstract}

\section{Introduction}

\begin{figure}[t!]
    \centering
    \includegraphics[width=\columnwidth]{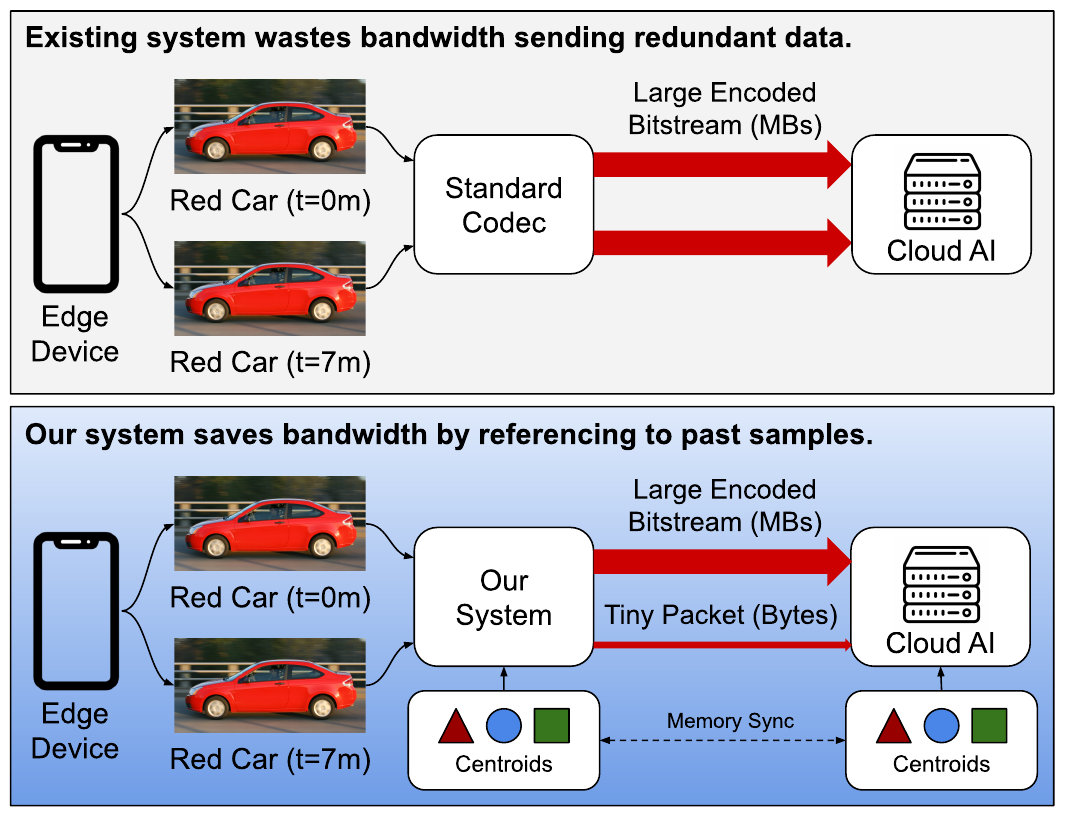}
    \caption{Rather than sending the full sensor measurement, the sender transmits only an index of a vector table. The receiver uses this index to retrieve the corresponding entry for downstream tasks.}
    \label{fig:teaser}
\end{figure}

Modern wireless sensing systems collect vast amounts of data to monitor and understand the physical world, enabling applications such as augmented and virtual reality, autonomous systems, and large-scale environmental surveys~\cite{mayton2017networked,vasisht2017farmbeats,chen2025yoro,
lee2025perception}. Their measurements range from scalar readings to high-dimensional modalities such as images, video, LiDAR, radar, and audio. Although sensing and local storage have become increasingly capable, moving these data from resource-constrained devices to a gateway or the cloud remains difficult, particularly over low-rate wireless links~\cite{yang2020insectrobot,naderiparizi2015wispcam,whisper_nsdi,vasisht2017farmbeats}. As sensing platforms become more capable, communication increasingly becomes the bottleneck in supporting downstream applications.


Traditional compression reduces communication by encoding observations compactly while preserving signal fidelity~\cite{wallace1991jpeg,h264,hvec}. However, downstream tasks may not require every detail of the original measurement. Semantic and task-oriented communication exploit this flexibility by transmitting information relevant to the receiver's task, potentially reducing bandwidth beyond what fidelity-oriented compression achieves~\cite{lee2023deepjointsourcechannelcoding,bourtsoulatze2019deepjscc,chen2024semanticunderwater,yang2024sg2sc}. However, these methods are typically optimized for fixed tasks, models, and training distributions. When the deployed data stream changes, maintaining performance often requires expensive retraining or online learning. Approaches that rely on large models also face deployment challenges on sensing devices with limited compute and on-chip memory~\cite{hojjat2024limitnet,huang2022real}.

\begin{figure}
    \centering
    \includegraphics[width=\linewidth]{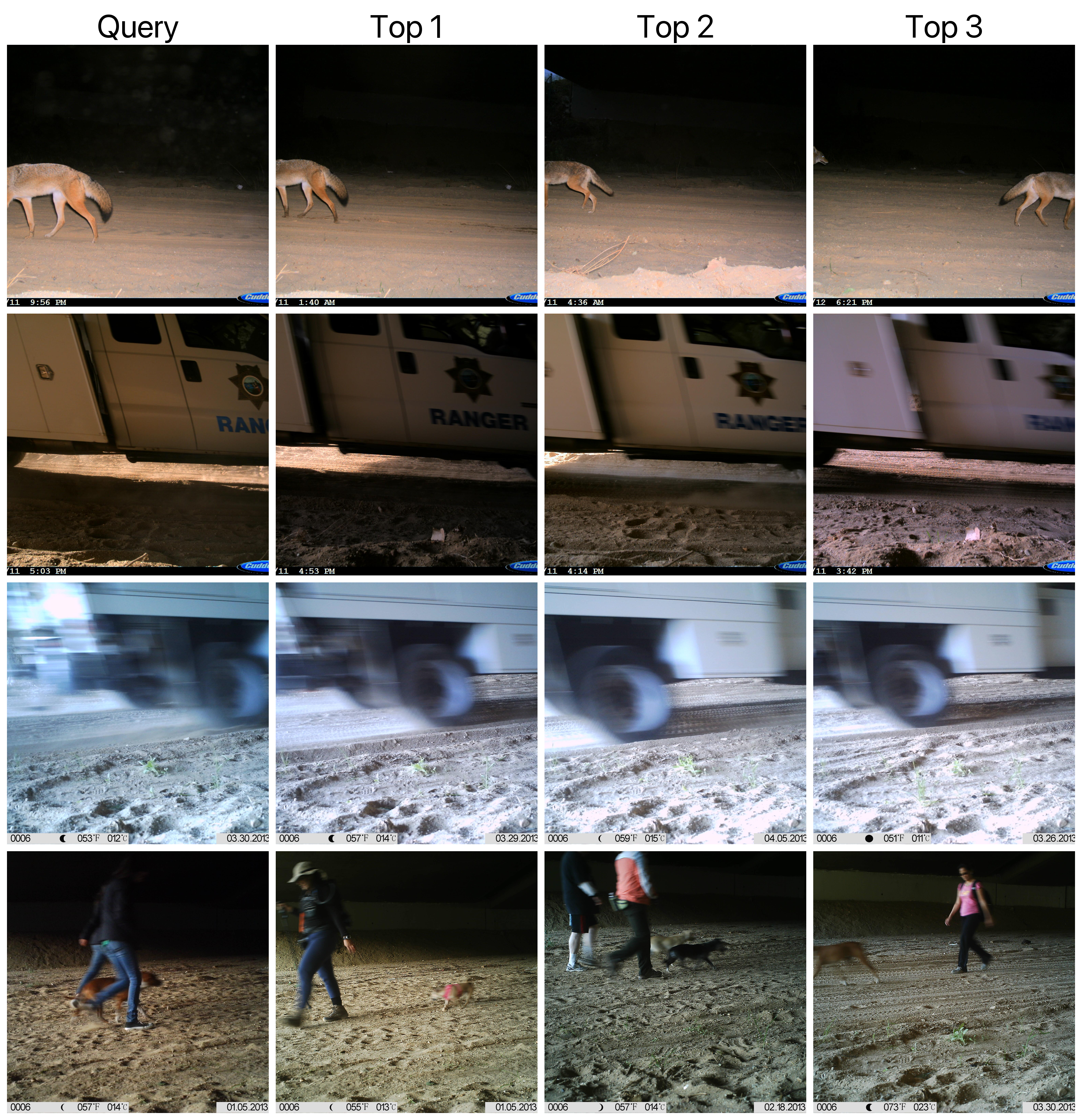}
    \caption{Example retrieval results from a vector table containing train examples from CCT. For each query image (left), we show the top-3 nearest neighbors retrieved from the table.}
    \label{fig:matches}
\end{figure}

\begin{figure*}[t]
    \centering
    \includegraphics[width=\linewidth]{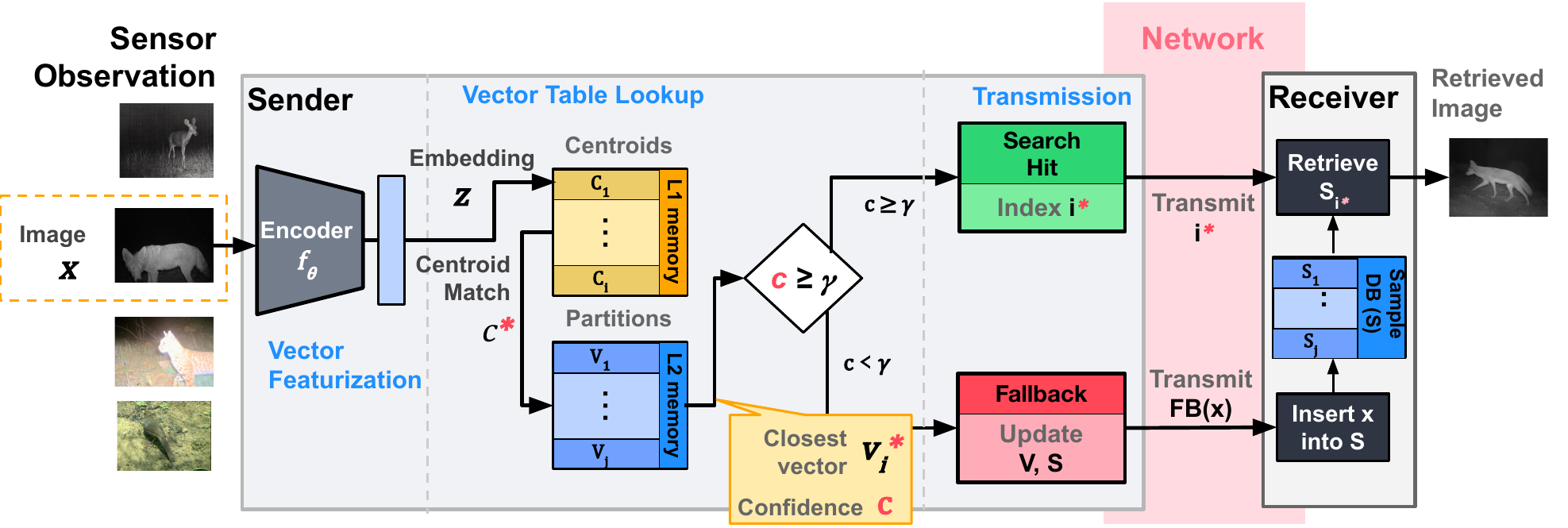}
    \caption{System overview of \tool{}. A sensor observation is encoded into a vector and used to query a table through L1 and L2 lookup. If the retrieved match exceeds a confidence threshold, the system transmits a reference to the stored sample. Otherwise, it falls back to transmitting the full observation and inserts the new sample into both the sender and receiver tables.}
    \label{fig:sys_overview}
\end{figure*}

This work is motivated by two simple observations about modern sensing systems. \textbf{First, the data transmitted by sensing systems is often highly redundant}. Sensor nodes often encounter similar objects, patterns, and contexts over time, even as their viewpoints or surroundings change. Over time, a large fraction of the transmitted data resembles content already observed by the same node. \textbf{Second, modern sensing platforms have access to abundant and inexpensive non-volatile memory}. This storage provides a persistent memory of examples that can grow as the environment changes.

Building on these insights, we propose \textbf{\tool}, a communication system that transmits data by \emph{reference} through vector retrieval. As shown in Figure~\ref{fig:teaser}, a conventional system transmits both images of a red car captured minutes apart, even though they contain similar information. \tool instead transmits the first image in full and sends a compact reference for the later, similar observation. \tool supports these references by maintaining a shared memory of previously transmitted samples at the sender and receiver. Figure~\ref{fig:matches} shows \tool's retrieval results on the Caltech Camera Trap (CCT) dataset~\cite{cct}. The retrieved samples preserve not only the key object but also finer details, such as a ranger truck, a person walking with a dog, or a single coyote at night, that coarse class labels cannot fully capture.


Many successful systems build on the idea of reusing previously transmitted information to reduce communication cost. For example, caches avoid repeatedly transferring identical content~\cite{berners1996hypertext}. Modern video codecs such as MPEG and H.264/H.265 reduce bitrate by encoding frames relative to previously transmitted reference frames~\cite{hvec}. Our work revisits this idea in the context of embedded sensing, where severe bandwidth and energy constraints make eliminating redundant transmissions especially powerful.


In this paper, we design and prototype \tool on an embedded platform comprising a microcontroller (MCU), SD-card storage, and Long Range (LoRa) communication. We also develop latency and energy models to identify the wireless links at which \tool becomes preferable to full transission. We evaluate \tool and five other baselines across four real-world datasets including image and radar workloads, where \tool{} traces the accuracy-payload size and accuracy-energy Pareto frontier.


\noindent\textbf{Contributions.} \begin{itemize}[leftmargin=10pt]
\item \textbf{\tool.} We introduce a communication paradigm that uses vector retrieval over persistent memory.

\item \textbf{Efficient retrieval algorithm.} We design a multi-level vector search structure that enables fast lookup over large tables stored in embedded non-volatile memory.

\item \textbf{Performance modeling.} 
We derive the link rate thresholds under which \tool becomes cheaper than full transmission in both latency and energy.

\item \textbf{Prototype and evaluation.} We implement \tool{} on a real MCU+SD+LoRa platform and show across four datasets that it achieves a stronger accuracy-communication and accuracy-energy Pareto frontier than baselines. 

\item \textbf{Online adaptation.} We demonstrate lightweight online adaptation under distribution shift.
Model-based baselines lose all accuracy while \tool{} retains classification performance by growing its table without encoder retraining.

\end{itemize}
\section{\tool}
\label{sec:method}


Figure~\ref{fig:sys_overview} illustrates the architecture of \tool. Instead of transmitting high-dimensional raw signals for reconstruction, the sender and receiver share a feature embedding space, where geometric proximity corresponds to feature similarity. At the sender, each observation is mapped into this space and matched against a vector table.
When a close match is found, the sender transmits only the index of the matched entry. When there is no close match, the sender transmits a fallback payload (e.g., a compressed sample) and inserts the new sample into the table, allowing future references. This design is modality-agnostic and applies to any sensor stream for which a shared similarity space can be defined. We next describe how \tool{} constructs and maintains the shared table, supports efficient retrieval, and handles reference and fallback transmissions.



\subsection{Vector Featurization}
\label{subsec:featurization}

A pre-trained, frozen encoder \(f_\theta\) is used for each sensing modality, ensuring that both endpoints operate in the same feature space. Given an observation \(x\in\mathcal{X}\), the sender computes a unit-norm embedding
\begin{equation}
    z \;=\; \frac{f_\theta(x)}{\|f_\theta(x)\|_2} \;\in\; \mathbb{R}^d.
\end{equation}
optionally followed by a random projector \(P\) to reduce the embedding dimension. Similarity between two vectors is measured with cosine,
\begin{equation}
    \mathrm{sim}(z_i,z_j)=z_i^\top z_j=\cos(\theta_{ij}).
\end{equation}

\subsection{Table Construction}
\label{subsec:table}

\tool{} maintains two synchronized tables that share a common index space. At the transmitter, a \emph{vector table} stores vector embeddings,
\begin{equation}
\mathcal{V}\;=\;\{v_i\}_{i=1}^{M}, \qquad v_i\in\mathbb{R}^d,
\end{equation}
while at the receiver, a \emph{sample table} stores the corresponding sensor payloads and optionally labels,
\begin{equation}
\mathcal{S}\;=\;\{(\pi_i,\,y_i)\}_{i=1}^{M}.
\end{equation}
The index $i$ is shared across $\mathcal{V}$ and $\mathcal{S}$. Transmitting an index \(i^\star\) allows the receiver to dereference \(\pi_{i^\star}\) (and \(y_{i^\star}\)) without reconstructing the raw observation through the channel.

At deployment, the tables can start either \textit{cold}, with both tables (near-) empty (\(M{\approx}0\)), or \textit{warm}, with the tables pre-populated using a prior over the data stream.



\subsection{Optimized Vector Table}
\label{subsec:ivf}

Direct nearest-neighbor search over the entire vector table \(\mathcal{V}\) becomes increasingly expensive as the table grows. A naive linear scan requires \(O(M \cdot d)\) operations per query, which quickly becomes impractical on resource-constrained devices. This cost is compounded by a tradeoff between memory speed and capacity, as memory that supports fast reads has low storage density and limited capacity, while higher-density memory offers greater capacity at the cost of higher access latency. As a result, a large vector table cannot fit entirely in fast memory, yet scanning it in slower memory incurs substantial access overhead. To address this tradeoff, we organize the vector table using an \textit{inverted file (IVF)} structure~\cite{jegou2010product}. IVF groups vectors into clusters and performs a two-pass search, first comparing the query against cluster centroids and then searching only the vectors in the selected cluster.

We store the small centroid table in fast Level-1 memory and the full vector table in larger, slower Level-2 memory. Let
\begin{equation}
\mathcal{C} = \{c_j\}_{j=1}^{K}, \qquad c_j \in \mathbb{R}^d
\end{equation}

denote the centroid table. Each centroid corresponds to a partition (or \textit{inverted list}), giving
\begin{equation}
\mathcal{V} = \bigcup_{j=1}^{K} \mathcal{V}_j,
\end{equation}
where \(\mathcal{V}_j\) contains all vectors assigned to centroid \(c_j\). The vectors within each partition are stored sequentially in Level-2 memory.

\prh{Query routing.} Given a query embedding \(z\), \tool{} performs a coarse search over the Level-1 centroid table in the first pass,
\begin{equation}
j^\star = \arg\max_{j \in {1,\dots,K}} z^\top c_j .
\label{eq:centroid}
\end{equation}
The selected centroid \(j^\star\) determines which partition \(\mathcal{V}_{j^\star}\) to probe in Level-2 memory.

In the second pass, \tool{} performs a fine-grained similarity search within the selected partition,
\begin{equation}
i^\star = \arg\max_{v_i \in \mathcal{V}_{j^\star}} z^\top v_i.
\label{eq:nn_search}
\end{equation}
For approximately balanced partitions, this reduces the number of vectors examined in Level-2 memory from \(M\) to approximately \(M/K\), in addition to the \(K\) centroid comparisons in Level-1 memory. This organization reduces both vector comparisons and accesses to slower memory.

\prh{Insertion and growth.} When a new vector \(z\) is inserted into the table (e.g., after a fallback transmission), it is assigned to the centroid selected by Equation~\ref{eq:centroid} and appended to the corresponding partition. To bound the number of vectors scanned in the second pass, each partition is assigned a maximum capacity \(C_{\max}\). When inserting a vector causes a partition \(\mathcal{V}_j\) to exceed this limit, the partition is split into two smaller partitions.

Specifically, the vectors in \(\mathcal{V}_j\) are clustered into two groups, producing two new centroids \(c_{j_1}\) and \(c_{j_2}\). The original centroid \(c_j\) is replaced by these two centroids in the Level-1 centroid table, and the vectors are redistributed into two corresponding partitions,
\begin{equation}
\mathcal{V}*j ;\rightarrow; \mathcal{V}*{j_1} \cup \mathcal{V}_{j_2}.
\end{equation}

\prh{More complex vector tables.} Depending on the task complexity and the memory system, we can also implement the vector table using more complex data structures that can provide higher recall.
One prominent category is graph-based approximate nearest-neighbor search indexes, such as Hierarchical Navigable Small World (HNSW)~\cite{hnsw} and Navigating Spread-out Graph (NSG)~\cite{nsg}.
These methods represent vectors as nodes connected by a sparse set of edges and search the graph by moving toward vectors that are more similar to the query.
They can offer higher recall than IVF, although the memory-access cost depends on the index configuration and search settings.

\subsection{Retrieval and Transmission}
\label{subsec:retrieval}
At runtime, the sender embeds a new observation \(x\) and performs nearest-neighbor search over \(\mathcal{V}\) using Equation~\ref{eq:nn_search}. A confidence score is computed either as \(c=z^\top v_{i^\star}\) or as a margin \(c=z^\top v_{i^\star}-z^\top v_{i^{(2)}}\) with \(i^{(2)}\) the second-nearest neighbor.

\prh{Search hit and fallback.} With a threshold \(\gamma\) tuned on an unlabeled validation stream to target a desired hit rate \(p_{\mathrm{hit}}\), the sender transmits the index \(i^\star\) when \(c\ge\gamma\) and a fallback payload \(\mathrm{FB}(x)\) when \(c<\gamma\), in which case $z$ and \((\pi,y)\) are appended to \(\mathcal{V}\) and \(\mathcal{S}\), respectively, at the specified index.

The communication cost follows from mixing index-only and fallback transmissions. Let \(B_{\mathrm{idx}}=\lceil\log_2 M\rceil + B_{\mathrm{hdr}}\) denote the bits for the index and protocol header/CRC, and let \(B_{\mathrm{fb}}\) be the measured fallback size. The expected bits per query at operating point \(\gamma\) are
\begin{equation}
\label{eq:expected-bits}
\mathbb{E}[\mathrm{bits}](\gamma) = p_{\mathrm{hit}}(\gamma) \cdot B_{\mathrm{idx}} + \bigl(1-p_{\mathrm{hit}}(\gamma)\bigr)\,B_{\mathrm{fb}}.
\end{equation}
At link rate \(R\) (kbps), the expected on-wire latency is \(1000\cdot \mathbb{E}[\mathrm{bits}](\gamma)/R\) ms.

For downstream tasks such as classification, let $h_i(\gamma)\in\{0,1\}$ indicate whether query $i$ results in a hit, let $\hat{y}_i^{\mathrm{local}}$ denote the prediction produced by the local retrieval path, let $\hat{y}_i^{\mathrm{cloud}}$ denote the prediction produced by the cloud model, and let $y_i$ be the ground-truth label. The effective prediction for query $i$ is
\begin{equation}
\hat{y}_i(\gamma)=
\begin{cases}
\hat{y}_i^{\mathrm{local}}, & \text{if } h_i(\gamma)=1,\\
\hat{y}_i^{\mathrm{cloud}}, & \text{otherwise.}
\end{cases}
\end{equation}
The resulting top-1 accuracy over $N$ queries is
\begin{equation}
\mathrm{Acc}(\gamma)
=
\frac{1}{N}\sum_{i=1}^{N}
\mathbf{1}\!\left\{\hat{y}_i(\gamma)=y_i\right\}.
\label{eq:accuracy}
\end{equation}
\section{When Does \tool Pay Off?}
\label{sec:model}

\tool{} trades local computation and storage access for shorter transmissions. We model query latency and energy to identify when this tradeoff pays off. We use lowercase \(t,e\) for individual operation or per-byte costs, and uppercase \(T,E\) for aggregate costs. Hardware-dependent inputs are shown in blue. Let $p=p_{\mathrm{hit}}(\gamma)$ denote the hit rate at a threshold that meets the application's quality requirement. \hw{Blue} denotes hardware-dependent variables.

\subsection{Query Cost Model}
\paragraph{Latency.}
Each query encodes an observation, looks up its embedding, and delivers a reference or fallback. Expected inference latency $T_{\mathrm{inf}}$ is the sum of local compute time $T_{\mathrm{local}}$ and radio transmission time $T_{\mathrm{tx}}$:
\begin{equation}
T_{\mathrm{inf}}=T_{\mathrm{local}}+T_{\mathrm{tx}}.
\label{eq:tinf}
\end{equation}
$s_{\mathrm{ref}}$ and $s_{\mathrm{fb}}$ are the sizes of a reference packet and a fallback packet, respectively. With delivery time $\tau(s)$ for $s$ bytes,
\begin{equation}
T_{\mathrm{tx}}=p \cdot \tau(s_{\mathrm{ref}})+(1-p) \cdot \tau(s_{\mathrm{fb}}).
\label{eq:txtotal}
\end{equation}
We approximate delivery at a constant rate $\hw{R}$ in bit/s, giving $\tau(s)=8 \cdot s/\hw{R}$.

Local computation includes table lookup, whose duration is denoted as $T_{\mathrm{lookup}}$. Encoding and table lookup occur on every query, while fallback preparation and table updates occur only on misses. Their expected local cost is
\begin{equation}
T_{\mathrm{local}}=\hw{t_{\mathrm{enc}}}+T_{\mathrm{lookup}}
+(1-p) \cdot (t_{\mathrm{fb}}+t_u),
\label{eq:tlocal}
\end{equation}
where $\hw{t_{\mathrm{enc}}}$, $t_{\mathrm{fb}}$, and $t_u$ are encoding time, fallback processing time, and vector table update time, respectively. For the IVF structure described above, lookup time comprises Level-1 memory read time $T_1$, Level-2 memory read time $T_2$, and distance computation time $T_d$. Lookup reads $K$ centroids and $m$ candidate vectors from the selected partition, each of size $s_e$ bytes. With read times per byte $\hw{t_{r,1}}$ and $\hw{t_{r,2}}$ and similarity time per stored byte $\hw{t_d}$,
\begin{equation}
\begin{aligned}
T_1 &= K \cdot s_e \cdot \hw{t_{r,1}},
& T_2 &= m \cdot s_e \cdot \hw{t_{r,2}}, \\
T_d &= (K+m) \cdot s_e \cdot \hw{t_d},
& T_{\mathrm{lookup}} &= T_1+T_2+T_d.
\end{aligned}
\label{eq:tlookup_ivf}
\end{equation}

\paragraph{Energy.}
Similarly, total inference energy $E_{\mathrm{inf}}$ is composed of encoding energy $\hw{e_{\mathrm{enc}}}$, lookup energy $E_{\mathrm{lookup}}$, communication energy $E_{\mathrm{comm}}$, fallback processing energy $e_{\mathrm{fb}}$, and table update energy $e_u$:
\begin{equation}
\begin{split}
E_{\mathrm{inf}}=
&\hw{e_{\mathrm{enc}}}+E_{\mathrm{lookup}}+E_{\mathrm{comm}}\\
&+(1-p) \cdot (e_{\mathrm{fb}}+e_u),
\end{split}
\label{eq:einf}
\end{equation}
With delivery energy $e(s)$ for $s$ bytes, transmission energy is
\begin{equation}
E_{\mathrm{comm}}=p \cdot e(s_{\mathrm{ref}})+(1-p) \cdot e(s_{\mathrm{fb}}).
\label{eq:etx}
\end{equation}
Under continuous transmission, $e(s)=(\hw{P_{\mathrm{tx}}}+\hw{P_{\mathrm{idle}}}) \cdot \tau(s)$, where $\hw{P_{\mathrm{tx}}}$ and $\hw{P_{\mathrm{idle}}}$ are radio transmission power and processor's idle power (optionally, in sleep mode), respectively.

Lookup reuses the stage times above. With total sender powers $\hw{P_1}$, $\hw{P_2}$, and $\hw{P_d}$ during the two reads and distance computation,
\begin{equation}
E_{\mathrm{lookup}}=\hw{P_1} \cdot T_1+\hw{P_2} \cdot T_2+\hw{P_d} \cdot T_d.
\label{eq:elookup_ivf}
\end{equation}
These powers include the processor and active memory or storage. Unused components are off. Update energy $e_u$ includes insertion and amortized maintenance, even when deferred. A frozen table has $e_u=t_u=0$. For varying payloads, costs are computed per query before averaging.

\subsection{Break-Even Conditions}
Direct transmission prepares and sends the fallback on every query, costing $t_{\mathrm{fb}}+\tau(s_{\mathrm{fb}})$ in time and $e_{\mathrm{fb}}+e(s_{\mathrm{fb}})$ in energy. Retrieval pays off when expected savings from hits exceed the added encoding and lookup costs. For a frozen table and constant-rate transmission, equating the two costs gives the break-even throughputs $R_T^\star$ for time and $R_E^\star$ for energy:
\begin{equation}
\begin{aligned}
R_T^\star &= \frac{8 \cdot p \cdot (s_{\mathrm{fb}}-s_{\mathrm{ref}})}
{\hw{t_{\mathrm{enc}}}+T_{\mathrm{lookup}} - p \cdot t_{\mathrm{fb}}}, \\
R_E^\star &= \frac{8 \cdot p \cdot (\hw{P_{\mathrm{tx}}}+\hw{P_{\mathrm{idle}}}) \cdot (s_{\mathrm{fb}}-s_{\mathrm{ref}})}
{\hw{e_{\mathrm{enc}}}+E_{\mathrm{lookup}} - p \cdot e_{\mathrm{fb}}}.
\end{aligned}
\label{eq:rate-break-even}
\end{equation}
Both $R_T^\star$ and $R_E^\star$ are link rates in bit/s. The numerators capture the expected bits avoided per query, weighted by the sender's total power during transmission for the energy threshold. Each denominator represents the added encoding and lookup cost, offset by the fallback preparation cost avoided on hits. For positive denominators, retrieval reduces latency when $R<R_T^\star$ and energy when $R<R_E^\star$. Below these rates, the savings from shorter transmissions outweigh the added local costs.

\subsection{Using the Model to Analyze Tradeoffs}
\begin{figure}[t]
\centering
\includegraphics[width=\columnwidth]{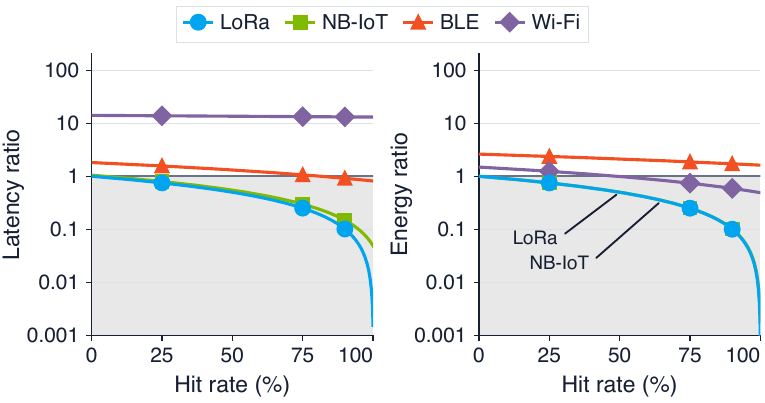}
\caption{Predicted latency and energy relative to direct transmission as the hit rate varies. Each wireless link uses its own throughput and radio power. \textbf{The shaded region indicates where \tool outperforms full data transmission.} Markers highlight hit rates of 25\%, 75\%, and 90\%. In the energy plot, the LoRa and NB-IoT curves nearly overlap.}
\label{fig:model-throughput}
\end{figure}

We now ask which wireless links make retrieval worthwhile, and whether latency and energy favor the same choice. We select Apollo4 Plus MCU and the Semtech LR2021 LoRa board from Section~\ref{sec:implementation} to measure model parameters listed in Table~\ref{tab:model-inputs}. 
Narrowband IoT (NB-IoT), Bluetooth Low Energy (BLE), and Wi-Fi 
use documented data rates and power consumption~\cite{nordic_nb_rate,nordic_nb_current,nordic_ble_radio,ti_cc3120}. We substitute these airtimes into Equations~\eqref{eq:txtotal} and~\eqref{eq:etx} to compute their cost ratios. These examples omit connection setup, receive windows, and retries. We use these parameters in the latency and energy equations to generate Figure~\ref{fig:model-throughput}, varying the hit rate.


\begin{table}[t]
\centering
\caption{Model inputs. Blue denotes hardware-dependent variables.
M: measured; S: selected; D: documented.
Radio cases are illustrative continuous-transfer approximations. $^\dagger$We send full sample, so there is no fallback preparation cost.}
\label{tab:model-inputs}
\small
\setlength{\tabcolsep}{3pt}

\begin{tabular*}{\linewidth}
{@{\extracolsep{\fill}}llc@{}}
\toprule
\textbf{Variable} & \textbf{Value} & \textbf{Src} \\
\midrule
\hw{$t_{\mathrm{enc}},\,e_{\mathrm{enc}}$}
  & 0.29 s, 3.40 mJ & M \\
\hw{$t_{r,1},\,t_{r,2},\,t_d$}
  & 381.47, 1162.72, 33.57 ns/B & M \\
\hw{$P_1,\,P_2,\,P_d$}
  & 10, 175, 10 mW & M \\
\hw{$P_{\mathrm{idle}}$}
  & 0.84 mW & M \\
\addlinespace[2pt]
$t_{\mathrm{fb}},\,e_{\mathrm{fb}}$
  & 0 s$^\dagger$, 0 J$^\dagger$ & S \\
$s_{\mathrm{ref}},\,s_{\mathrm{fb}}$
  & 6 B, 50 KiB & S \\
$K,\,m,\,s_e$
  & 64, 256, 128 B & S \\
\end{tabular*}
\par\nointerlineskip
\begin{tabular*}{\linewidth}
{@{\extracolsep{\fill}}lrrc@{}}
\midrule
\textbf{Link} & \textbf{\hw{$R$} (Rate)} & \textbf{$\hw{P_\mathrm{{tx}}}$ (TX power)}
  & \textbf{Src} \\
\midrule
LoRa   & 1.76 kbit/s & 86.44 mW & M \\
NB-IoT & 60 kbit/s   & 851 mW   & D \\
BLE    & 1 Mbit/s    & 14.4 mW  & D \\
Wi-Fi  & 16 Mbit/s   & 802.8 mW & D \\
\bottomrule
\end{tabular*}
\end{table}

Figure~\ref{fig:model-throughput} applies the break-even conditions in Equation~\eqref{eq:rate-break-even} by varying the hit rate while keeping each link's throughput and radio power fixed per curve. Each plot shows \tool's cost divided by the cost of direct transmission. A ratio of one marks the break-even point. Below one, \tool is faster in the latency panel or uses less energy in the energy panel. For all four links, these shaded regions correspond to $R<R_T^\star(p)$ and $R<R_E^\star(p)$, respectively.

\prh{Retrieval pays off on low-throughput links even at modest hit rates.} On low-throughput links such as the LoRa and NB-IoT examples, replacing a full transmission with a short reference saves enough time and energy to offset encoding and lookup. At a 25\% hit rate, \tool reduces latency by approximately 25\% for LoRa and 20\% for NB-IoT, and reduces energy by approximately 25\% for both. High hit rates are therefore not required for retrieval to help when transmission is expensive.

\prh{Higher hit rates amplify the savings.} As the hit rate rises, more queries send short references instead of full observations. Equation~\eqref{eq:rate-break-even} captures this effect through the expected transmission savings, which grow with $p$. At a 90\% hit rate, \tool reduces latency by approximately 90\% for LoRa and 85\% for NB-IoT, and reduces energy by approximately 90\% for both. This corresponds to about $9.8\times$ and $6.7\times$ speedups, respectively. When transmission dominates the total cost, the remaining cost is approximately proportional to the miss rate, $1-p$, so frequently reusing references yields large gains.


\prh{On high-throughput links, \tool{} may not improve both latency and energy.} Higher throughput leaves less transmission time for retrieval to save, while energy savings also depend on the sender's power consumption during transmission. BLE reduces latency only above an 81.6\% hit rate and does not save energy under these inputs. Wi-Fi saves energy above a 49.2\% hit rate but increases latency even at a perfect hit rate. Thus, \tool is most attractive when full transmissions are expensive and the workload frequently admits references at the required quality.
\section{Implementation}
\label{sec:implementation}


We describe the encoder training pipeline and the prototype implemented for deployment.


\prh{Encoder training.} Our implementation supports image and radar inputs, using a lightweight encoder for each modality. We train these encoders through knowledge distillation. For image inputs, the teacher is a pretrained DINOv2 ViT-S/14 model~\cite{oquab2023dinov2} that produces 384-dimensional L2-normalized embeddings, and the student is MCUNet~\cite{lin2020mcunet}, producing 128-dimensional embeddings. For radar inputs, we use ConvNeXt-Tiny~\cite{liu2022convnet} as the teacher and RadarTCN~\cite{li2024radartcn} as the student, with the student producing 64-dimensional embeddings. All training is performed on a single NVIDIA RTX 4090 GPU.

\prh{MCU prototype.} We implement the on-device prototype on an Ambiq Apollo4 Plus evaluation board~\cite{ambiq} running at 192~MHz. The neural network is executed with TensorFlow Lite Micro (TFLM), while the vector database is stored on a removable microSD card~\cite{adafruit}. In our prototype, the database stores 50,000 vectors of dimension 128 with 512 coarse clusters. The neural network uses a 600~kB tensor buffer in SRAM.

The MCU uses a two-tier memory layout. At startup, it loads the IVF routing metadata into SRAM, including the centroid table and partition lengths. The full database vectors and labels remain on the SD card. Given an input image, the MCU first runs the TFLM model to produce an embedding, then finds the nearest centroid in SRAM, and finally reads only the selected partition from the SD card for flat search within that partition. Each partition is stored as a contiguous binary file, so online updates can occur per file without affecting other partitions.

With this configuration, centroid lookup and distance computation in SRAM take 3.4~ms, reading the selected partition from the microSD card takes 38.1~ms, and computing distances within that partition takes 1.1~ms, averaged across 10 measurements. SD card access accounts for most of the measured retrieval latency.

Wireless communication uses a Semtech LR2021 sub-GHz transceiver, a LoRa 
radio commonly deployed in IoT sensing networks. The radio is a Wio-LR2021 module on a Semtech LR2021 LoRa Plus expansion board. We transmit at 868.1~MHz with a 125~kHz bandwidth, spreading factor~9, and coding rate $4/5$. The payload is capped at 255~B and larger messages are fragmented.


\begin{figure*}[t!]
    \centering
    \includegraphics[width=\textwidth]{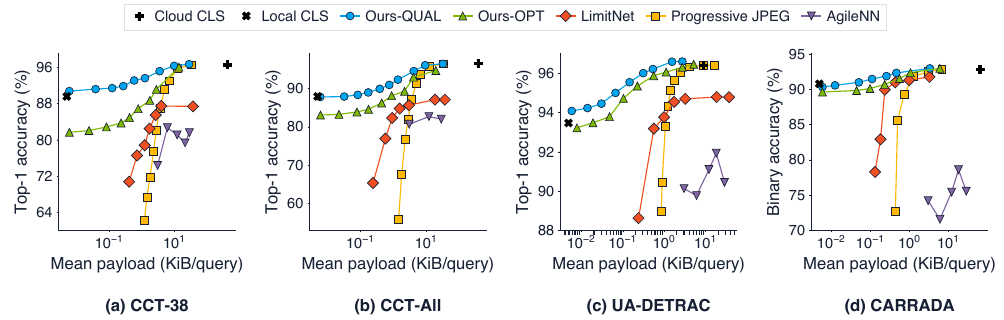}
    \caption{\textbf{Accuracy-communication tradeoff across four workloads.} Application accuracy versus mean transmitted payload (log scale) for four datasets. Accuracy is top-1 for the first three workloads and binary for (d), which is the mean agreement over four presence bits. \method{CloudCLS} (\ding{58}) and \method{LocalCLS} (\ding{54}) are single points. \method{Ours-QUAL} traces the Pareto frontier.}
    \label{fig:accuracy-comm-pareto}
\end{figure*}

\section{Evaluation}
\label{sec:evaluation}


We evaluate communication by retrieval on three image and one radar datasets that represent real-world sensing workloads. We present: 
(1) the accuracy-bandwidth Pareto frontier, (2) latency and energy analysis on an MCU, (3) online adaptation, and lastly, (4) a qualitative analysis of the transmitted packets.

\subsection{Experimental Setup}
\label{subsec:setup}

\noindent\textbf{Datasets.} We evaluate our system and baselines on real-world image and radar workload:
\begin{itemize}[leftmargin=10pt]
    
    
    \item \textbf{CCT-38} is a subset of the Caltech Camera Trap (CCT) dataset~\cite{cct}, which contains wildlife images for species classification across 20 classes. CCT-38 uses images from a single camera location (38), representing a static deployment that repeatedly observes the same environment.

    \item \textbf{CCT-All} contains images from all 140 camera locations in the full CCT dataset. 
    
    \item \textbf{UA-DETRAC}~\cite{uadetrac} contains real-world traffic video from stationary cameras, collected for vehicle detection and tracking. We crop each frame to the largest annotated vehicle and classify it into four vehicle classes.

    \item \textbf{CARRADA}~\cite{carrada} pairs camera images with radar range-Doppler maps. We use the radar maps for four-bit presence detection in autonomous vehicle perception, with one bit each indicating an empty scene or the presence of a pedestrian, a cyclist, or a car.

\end{itemize}




\prh{Baselines.} We compare our system against representative approaches. Local and cloud classifiers represent the two extremes, where the former sends only a class label and the latter sends the full image. Other approaches lie in the middle, using onboard computation or compression to reduce communication payload size.

\begin{itemize}[leftmargin=10pt]

\item \textbf{Local classifier} performs full on-device classification, and the predicted label is transmitted. We choose MCUNet~\cite{lin2020mcunet} for the image workload and RadarTCN~\cite{li2024radartcn} for the radar workload.

\item \textbf{Cloud classifier} receives full-resolution samples transmitted directly without local processing. It classifies images using ViT-Large~\cite{dosovitskiy2021an} and radar samples using ConvNeXt-Tiny~\cite{liu2022convnet}, fine-tuned on the corresponding dataset.


\item \textbf{Progressive JPEG~\cite{wallace1991jpeg}} is a compression baseline that progressively refines the reconstructed image as more bytes arrive at the receiver. The reconstructed image is classified using the cloud classifier.

\item \textbf{LimitNet~\cite{hojjat2024limitnet}} transmits a compressed feature map, which the receiver uses to reconstruct the image for classification.

\item \textbf{AgileNN~\cite{huang2022real}} transmits local class scores and a compressed intermediate feature map to the cloud. AgileNN's own remote classifier processes the reconstructed features, and its scores are combined with the local scores for the final prediction.

\end{itemize}

\prh{Our models.}
We evaluate two variants of \tool against these baselines: \method{OPT} and \method{QUAL}. \method{OPT} uses a smaller backbone and is optimized for resource efficiency. \method{QUAL} uses a larger backbone encoder, the largest that fits on the MCU, and is optimized for output quality. For image workloads, \method{OPT} uses mcunet-in1, while \method{QUAL} uses the larger mcunet-in2. For the radar workload, \method{OPT} uses a slimmed-down RadarTCN with fewer channels and residual blocks, whereas \method{QUAL} uses the full RadarTCN backbone. In most experiments, \short's vector table does not accept new observations. Online insertion is enabled only in Section~\ref{subsec:coldstart} to isolate the effect of online learning. Fallback payloads are classified using the cloud classifier.

\begin{figure*}[t!]
    \centering
    \includegraphics[width=\textwidth]{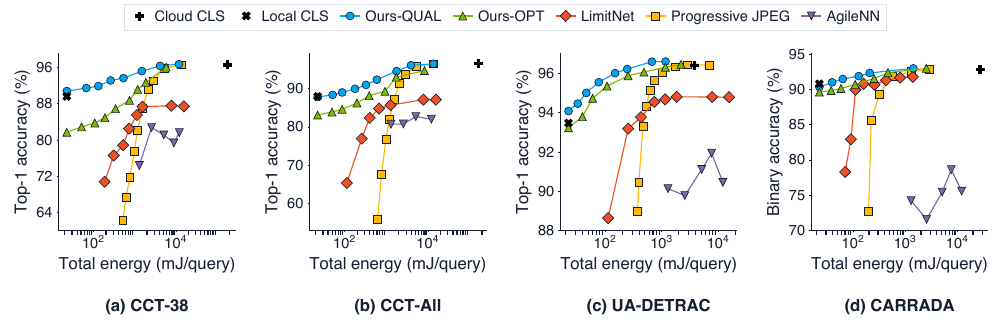}
    \caption{\textbf{Accuracy-energy tradeoff across four workloads.} Application accuracy versus total energy (log scale) for four datasets. \method{CloudCLS} (\ding{58}) and \method{LocalCLS} (\ding{54}) are single points. \method{Ours-QUAL} traces the Pareto frontier.}
    \label{fig:accuracy-energy-pareto}
\end{figure*}

\prh{Metrics.}
We evaluate each system using both task accuracy and system efficiency metrics. For task performance, we report classification accuracy on the query set: top-1 accuracy on the three image workloads and binary accuracy on CARRADA. Binary accuracy here is the mean agreement between predicted and ground-truth labels over four presence bits.

For communication efficiency, we report the empirical average number of transmitted bits per query. We measure MCU latency using the cycle counter and energy using a Joulescope JS220~\cite{joulescope_js220}, averaging over 10 trials. We estimate radio costs using a hardware-calibrated model built from 40-60 frames at each of 11 payload sizes. For messages larger than 255 B, we sum the modeled costs of all fragments, including per-fragment overhead.

\subsection{Accuracy vs. Bandwidth Savings }
\label{subsec:pareto}
We sweep each method across its communication control parameter and plot accuracy against the mean transmitted payload per query, as shown in Figure~\ref{fig:accuracy-comm-pareto}. We report top-1 accuracy for CCT-38, CCT-All, and UA-DETRAC, and binary accuracy for CARRADA. For \short, varying $\gamma$ changes the expected hit probability $p_{\mathrm{hit}}$ in Equation~\eqref{eq:expected-bits}. A stricter threshold increases accuracy at the cost of higher communication. For \method{ProgJPEG}, \method{LimitNet}, and \method{AgileNN}, we vary JPEG quality, rate setting, and split ratio, respectively.

The split-inference baselines provide limited improvements over local classification in these experiments. \method{AgileNN} is dominated by local classification on all four datasets, incurring larger payloads without improving accuracy. \method{LimitNet}'s accuracy increases with payload before largely plateauing. Its best setting remains below local accuracy on both CCT datasets, by 1.98 and 0.82 percentage points, leaving a 9.02-point gap to cloud classification on CCT-38. On UA-DETRAC and CARRADA, its best settings improve over local accuracy by 1.35 and 0.98 points at 4.99 and 3.25~KiB/query, respectively.

\method{ProgJPEG} exhibits a sharp accuracy increase with payload and approaches cloud accuracy on all four datasets. It is therefore competitive when the accuracy-loss tolerance is small. On CCT-38 and CCT-All, it reaches within 0.07 and 0.12 percentage points of cloud accuracy with 23.8$\times$ and 12.0$\times$ payload reductions, respectively. However, its accuracy drops substantially at smaller budgets. It needs $1.22$ to $4.95$~KiB/query to match local classification on all datasets. At the leftmost point, \method{ProgJPEG} falls $4.5$ points below a local classifier on UA-DETRAC, $18.1$ on CARRADA, and $27.3$ and $31.9$ on CCT-38 and CCT-All.

\short contributes operating points to the observed Pareto frontier on every dataset, including budgets supporting one to three orders of magnitude less communication than cloud classification. On CCT-38, \method{OPT} reaches 96.04\% accuracy with a 30.7$\times$ payload reduction, only 0.59 percentage points below cloud accuracy. At a smaller budget, \method{QUAL} achieves 92.74\% with a 988$\times$ reduction, exceeding local accuracy by 3.15 points. On CCT-All, \method{OPT} achieves 94.63\% with a 19.6$\times$ reduction and 92.20\% with a 112.1$\times$ reduction. On UA-DETRAC, \method{QUAL} reaches 96.22\% with a 16.6$\times$ reduction, within 0.19 points of cloud accuracy. On CARRADA, \method{QUAL} achieves 92.59\% at 0.90~KiB/query, compared with \method{ProgJPEG}'s best accuracy of 92.80\% at 6.72~KiB/query.

\begin{figure}[h]
    \centering
    \includegraphics[width=.95\linewidth]{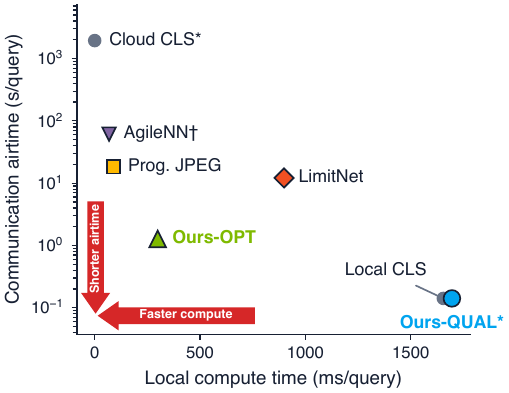}
    \caption{Local compute time and communication airtime measured on Apollo 4P MCU and LoRa (+14 dBm, 868.1~MHz, BW~125~kHz, SF~9), respectively. Methods are compared at matched accuracy, except higher-accuracy references (*) included for context and AgileNN ($\dagger$), which cannot reach the target. \textbf{\method{Ours-OPT} aims to reduce both times, thus achieving the lowest total time.}}
    \label{fig:localtime-commtime}
\end{figure}

\subsection{Latency Analysis}
\label{subsec:lat-and-energy}

Section~\ref{subsec:pareto} compares methodologies in terms of mean payload size. Here, we account for both communication airtime and local compute time. Reducing the transmitted payload may require substantial MCU computation, while lightweight local processing can leave communication as the dominant cost. \method{CloudCLS} and \method{LocalCLS} represent the two extremes. \method{CloudCLS} transmits the camera-generated source JPEG, averaging $382$~KiB and requiring approximately $1{,}945$~s of airtime. Because the JPEG is provided directly by the camera, \method{CloudCLS} incurs no additional local computation. \method{LocalCLS} performs the classification locally, requiring $1.65$~s of computation but only $0.141$~s of airtime to transmit its prediction.

The other baselines lie between these extremes. Progressive JPEG requires 89~ms of local computation but 18.3~s of airtime, while AgileNN requires 68~ms of computation and 61.3~s of airtime. Moreover, AgileNN does not reach the target accuracy. LimitNet reduces airtime to 12.1~s but increases local computation to 898~ms. Thus, although these methods reduce the transmitted payload, communication remains the dominant cost.

\tool's \method{OPT} achieves the best overall compute and communication trade-off, requiring 299~ms of local computation and 1.27~s of mean airtime. It sends a six-byte reference for 91.9\% of queries and uses the JPEG fallback for only the remaining 8.1\%. As a result, \method{OPT} achieves the lowest compute-plus-airtime cost among all methods at 1.57~s.

Notably, communication airtime spans more than four orders of magnitude, whereas local computation remains below 2~s per query for all methods. This asymmetry shows that a modest amount of additional local computation can substantially reduce communication time, affirming our conclusion in Section~\ref{sec:model}. 

\subsection{Energy Analysis}
\label{subsec:energy}

Figure~\ref{fig:accuracy-energy-pareto} plots application accuracy against total energy as we vary each method's communication control parameters.

\prh{\tool versus \method{LocalCLS}.} At the low-energy end, \method{LocalCLS} is competitive with the cheapest configuration of \tool. At \method{LocalCLS}'s energy consumption, \method{QUAL} shows comparable accuracy as \method{LocalCLS}. \method{OPT} has the lowest measured energy consumption on every dataset, using $1.5$ to $2.4\times$ less energy than \method{LocalCLS}, but at an accuracy cost of $0.25$ to $7.92$ points. 
However, \method{LocalCLS} only provides a single operating point. It saturates the MCU’s on-chip resources, and cannot accommodate a more accurate classifier without upgrading the hardware. In contrast, Recurio supports elastic operation, allowing the user to expend additional energy for higher application accuracy.

\prh{\tool versus \method{CloudCLS}.} At the high-energy end, \method{QUAL} matches \method{CloudCLS} accuracy on CCT-38, UA-DETRAC and CARRADA using $5.5$ to $27.1\times$ less energy, and falls within $0.14$ points on CCT-All at $14.7\times$ less energy.

\prh{\tool versus split-inference baselines.}
On CCT-38, UA-DETRAC, and CARRADA, \tool dominates every split-inference baseline configuration we measured.
On CCT-All, the same holds for all operating points of the baselines but the maximum-quality \method{ProgJPEG} setting, which is $0.02$ points more accurate than \method{QUAL}'s best setting.
At local accuracy, the lowest-energy \tool configuration reduces energy consumption by a factor of $7.4$ to $55.2$ relative to the lowest-energy competing baseline that reaches the same accuracy: \method{ProgJPEG} on the two CCT datasets and \method{LimitNet} on UA-DETRAC and CARRADA.
Furthermore, \tool offers operating points for different energy budgets on the Pareto frontier. These operating points allow users to trade additional energy for higher accuracy on the same MCU hardware.

\begin{figure}[h!]
    \centering
    \includegraphics[width=.83\columnwidth]{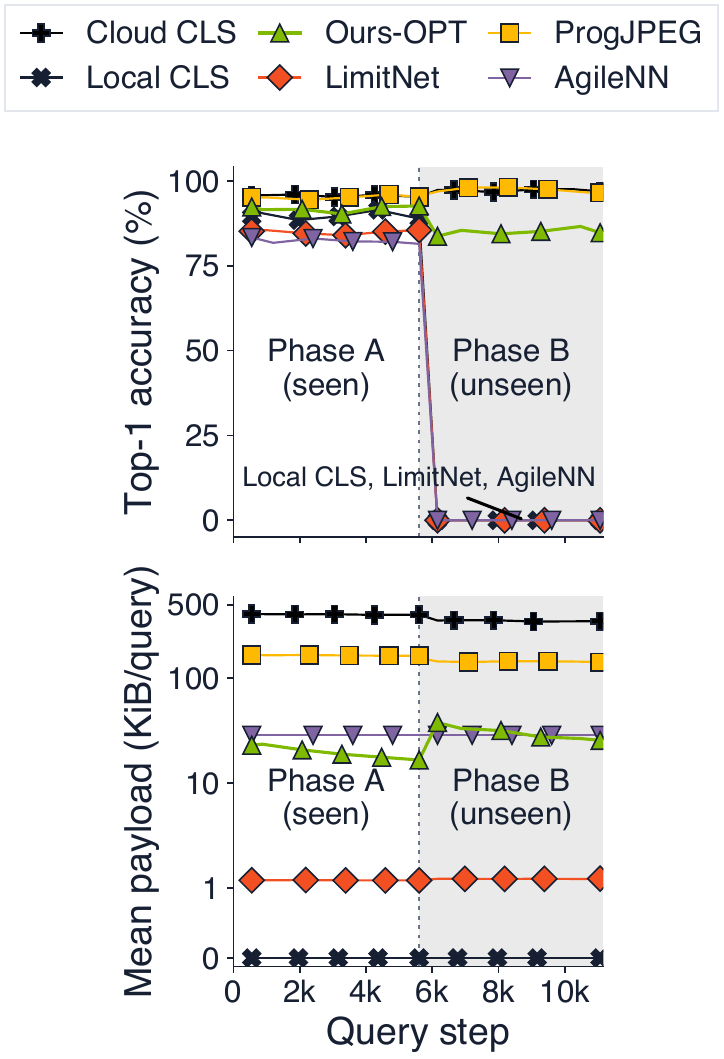}
    \caption{\textbf{Online adaptation under class shifts on the CCT-All dataset.} Top: classification accuracy. Bottom: mean communication payload. At step 5,604, the stream switches from seen classes to unseen classes (shaded). Curves show three-seed means over 553-query trailing windows, and sampled points are connected for clarity. The payload axis is linear below 1 KiB/query and logarithmic above.}
    \label{fig:online}
\end{figure}




\newlength{\fgw}\setlength{\fgw}{0.168\textwidth}
\newlength{\fgh}\setlength{\fgh}{0.75\fgw}      

\newcommand{\fgcell}[2]{%
  \begin{tabular}{@{}c@{}}%
    {\small #1}\\[1.5pt]%
    #2%
  \end{tabular}}
\newcommand{\fgtcell}[2]{%
  \begin{tabular}{@{}c@{}}%
    {\small #1}\\[1.5pt]%
    \parbox[c][\fgh][c]{1.7cm}{\centering #2}%
  \end{tabular}}
\newcommand{\fghead}[1]{%
  \begin{tabular}{@{}c@{}}\parbox[t]{\fgw}{\centering \textbf{#1}}\end{tabular}}

\begin{figure*}[t]
  \centering
  \setlength{\tabcolsep}{1.8pt}
  \renewcommand{\arraystretch}{1.08}
  \begin{tabular}{@{}cccccc@{}}
    \fghead{Query} &
    \fghead{\tool} &
    \begin{tabular}{@{}c@{}}\parbox[t]{1.7cm}{\centering \textbf{LocalCLS}}\end{tabular} &
    \fghead{ProgJPEG} &
    \fghead{AgileNN} &
    \fghead{LimitNet} \\[5pt]
    \fgcell{$172{,}176$ B}{\includegraphics[width=\fgw]{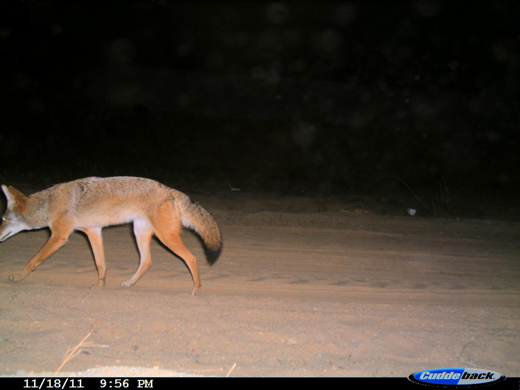}} &
    \fgcell{$6$ B}{\includegraphics[width=\fgw]{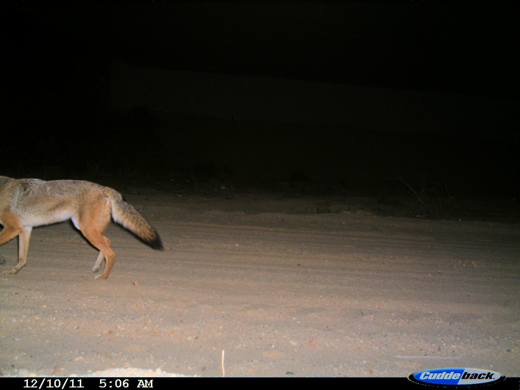}} &
    \fgtcell{$5$ B}{\textbf{``coyote''}} &
    \fgcell{$3{,}090$ B}{\includegraphics[width=\fgw]{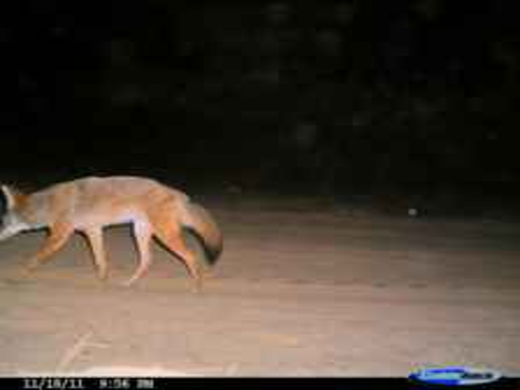}} &
    \fgcell{$12{,}312$ B}{\includegraphics[width=\fgw]{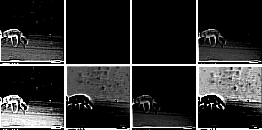}} &
    \fgcell{$5{,}268$ B}{\includegraphics[width=\fgw]{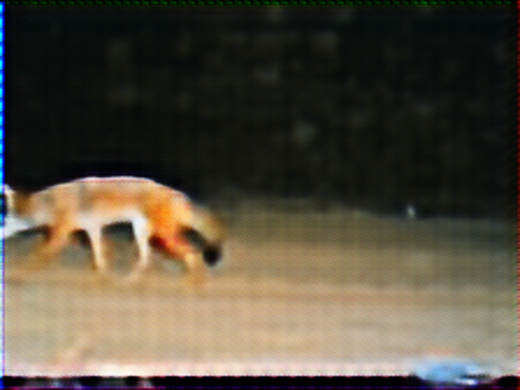}} \\[40pt]
    \fgcell{$191{,}668$ B}{\includegraphics[width=\fgw]{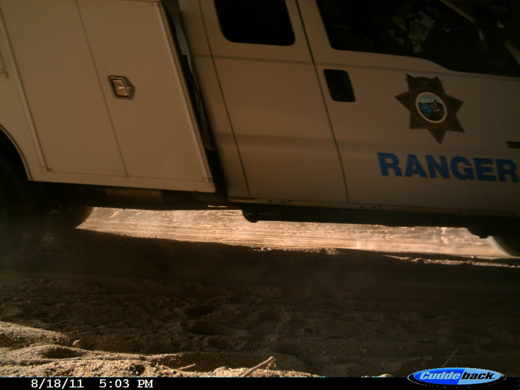}} &
    \fgcell{$6$ B}{\includegraphics[width=\fgw]{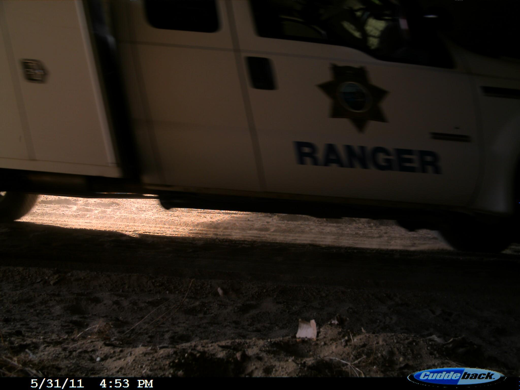}} &
    \fgtcell{$5$ B}{\textbf{``car''}} &
    \fgcell{$4{,}168$ B}{\includegraphics[width=\fgw]{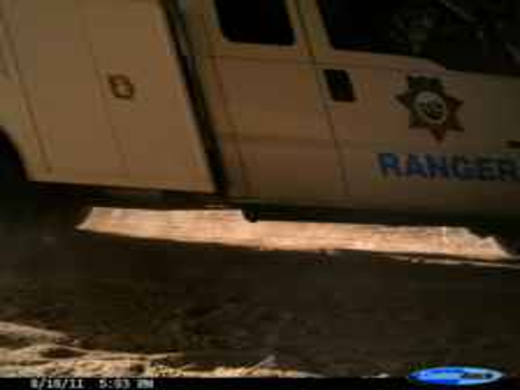}} &
    \fgcell{$12{,}312$ B}{\includegraphics[width=\fgw]{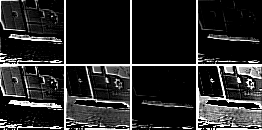}} &
    \fgcell{$5{,}620$ B}{\includegraphics[width=\fgw]{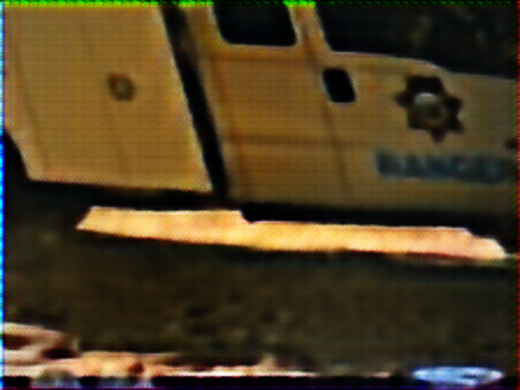}} \\[4pt]
  \end{tabular}
  \caption{\textbf{Qualitative comparison of the packet content that can be
  retrieved from the transfer between \tool and the baselines.} For each query
  image (left), we present the data retrieved from the transmitted packet of
  each method, with the packet size for that example above each cell.
  \textsc{AgileNN} does not reconstruct the input at the receiver side but
  instead feeds the transmitted feature maps into a classifier. \method{ProgJPEG} operates at $q{=}40$, \method{AgileNN} at split $0.67$, and \method{LimitNet} at full keep ratio with an eight-bit latent. Query and reference are $21$ and $79$ days apart, respectively.}
  \label{fig:qualitative}
\end{figure*}

\subsection{Online Adaptation}
\label{subsec:coldstart}

Real-world sensing workloads can encounter classes absent from the training data. For example, a camera trap may capture a species that its classifier was not trained to recognize. We use CCT-All to evaluate how \tool and the baselines respond to such class shifts. For this experiment only, \tool uses a growing vector table that appends unseen entries.

As shown in Figure~\ref{fig:online}, the stream is split into two phases (A $\to$ B) with disjoint class sets, with the vertical dotted line marking the shift boundary at step 5,604. The table is initialized with Phase-A samples, but not Phase-B samples. After the shift, all systems must process queries from unseen classes.

After the class shift, the local classifier, LimitNet, and AgileNN completely lose their ability to classify the incoming queries, dropping to 0\% top-1 accuracy. Cloud classification and Progressive JPEG maintain their accuracy but incur high communication costs by transmitting high-fidelity payloads. In contrast, \tool experiences a brief dip in accuracy and a modest increase in mean payload before quickly stabilizing at 83-87\% accuracy on previously unseen classes while transmitting 24-38~KiB per query.

\prh{System cost for \tool.} This adaptation incurs essentially no additional system overhead. The encoder remains fixed, and only the vector table grows. In contrast, adapting model-based baselines would require expensive offline retraining or on-device learning.




\subsection{Qualitative Analysis}

We qualitatively compare the content of the transmitted packets and the information retrievable at the receiver between \tool and the baseline methods.

Figure~\ref{fig:qualitative} shows what a receiver obtains from each method's packet on two CCT-All queries. On a search hit, \tool transmits a six-byte packet comprising a four-byte header and a two-byte reference index, and the receiver dereferences the index to an image it holds. 
\method{LocalCLS} transmits a five-byte packet comprising a header and a class index. 
The remaining baselines transmit an encoding of the observation. 
\method{ProgJPEG} operates at $q{=}40$, the cheapest setting that matches \tool's accuracy.  
\method{AgileNN} operates at split $0.67$, and \method{LimitNet} at full keep ratio with an eight-bit latent, both at their most accurate measured configurations. 

In the \tool hit cases, the reference image not only shows a subject of the same class, but also conveys more elaborate details in the query image using a six-byte packet. 
In the first query, \tool presents a reference image containing a single coyote passing the same site at night.
In the second query, the reference image shows a visually similar ranger truck to the query.
Neither pair comes from the same camera trigger event: query and reference are $21$ and $79$ days apart respectively. Although \method{LocalCLS} gets the correct prediction for both queries: ``coyote'' and ``car'', it omits such visual details.
For example, an authorized ranger truck and an unauthorized vehicle both reach the receiver as ``car'', and \method{LocalCLS} cannot restore the distinction.
\method{ProgJPEG} keeps the subject identifiable, but spends $3{,}090$~B on the coyote and $4{,}168$~B on the truck, $515$ and $695 \times$ \tool's six-byte packet. 
Similarly, \method{LimitNet} sends $5{,}268$~B on the coyote and $5{,}620$~B on the truck, $878$ and $937 \times$ \tool's packet. 
\method{AgileNN} sends a quantized feature tensor to a receiver-side classifier, spending $12{,}312$~B, $2{,}052 \times$ \tool's packet. 
The \method{AgileNN} architecture contains no decoder back to pixels, so it provides no visual detail for a human to audit.
Because \tool's index dereferences to a stored image, for six bytes the receiver can inspect high-resolution details that a class label or a compressed payload may lose.

\section{Related Work}


Our work is inspired by semantic communication and the recent surge in research toward knowledge-driven wireless communication networks~\cite{10554663}. There is a wealth of prior work spanning image compression, region-of-interest detection, and edge offloading~\cite{wallace1991jpeg,yan2017progressivejpeg,hojjat2024limitnet,lee2025hypercam}. Most closely related to our work are data-driven approaches that utilize deep learning to enhance the efficiency of wireless communication networks, particularly in scenarios involving a resource-constrained transmitter (e.g., low bandwidth, limited power, high packet loss) paired with a resource-abundant receiver.

Most relevant are AgileNN and LimitNet~\cite{hojjat2024limitnet,huang2022real}. LimitNet is an image compression and offloading model for extremely resource-constrained devices. Here, a lightweight content-aware progressive encoding scheme is implemented that prioritizes high-value data during transmission based on the content of an image. On the other hand, AgileNN is a neural network (NN) offloading technique that is capable of achieving real-time NN inference on embedded devices. Both works have demonstrated significant performance improvement in terms of latency and device resource utilization. In contrast, {\short} is able to outperform in terms of latency, energy-efficiency, and communication overhead across several datasets. 

Another related work is~\cite{lee2023deepjointsourcechannelcoding}, where an iterative source error correction (ISEC) decoding scheme for deep-learning-based joint source-channel coding (DeepJSCC) is proposed for more efficient encoding and decoding. This work demonstrates that ISEC improves distortion and perceptual quality metrics (compared to one-shot DeepJSCC) using image datasets.  In~\cite{hu2020starfish}, a DNN-based image compression framework is proposed for Internet-of-Things (IoT) cameras, demonstrating significant performance improvement in compression, latency, and energy efficiency compared to JPEG-based solutions. In~\cite{DeepCOD}, a deep compressive offloading framework is proposed to reduce end-to-end offloading latency without compromising accuracy. Here, an asymmetric encoder-decoder structure is used that integrates compressive sensing with deep neural networks. This work demonstrates that offloading latency can be significantly reduced with minimal accuracy drop for mobile-edge-network scenarios. Both methods are evaluated by baseline methods~\cite{hojjat2024limitnet,huang2022real} and are outperformed.

While all of these techniques offer many advancements and take a step towards enabling knowledge-driven communication, they still have limitations. Many approaches are trained on fixed datasets, rely on computationally expensive models, or lack adaptability to diverse deployment scenarios, thereby limiting their applicability. The proposed work aims to bridge the gap by introducing a lightweight and adaptive framework that takes advantage of the key observation that real-world sensor data has inherent redundancies. 



\section{Discussion}

\prh{When is \short effective?} \short performs best in sensing deployments where observations are temporally redundant, such as camera traps, environmental monitoring, or industrial inspection. In highly dynamic environments where observations rarely repeat, the system falls back to transmitting full samples more frequently, reducing communication savings but maintaining correctness through the fallback path.

\noindent\textbf{Vector table growth and storage.}
The vector table grows over time as new samples are inserted. However, modern IoT devices commonly include gigabytes of non-volatile storage, making large tables practical. Our evaluation shows that retrieval performance saturates with relatively modest table sizes, suggesting that extremely large tables are often unnecessary. In practice, deployments could bound table size using simple policies such as age-based eviction, clustering, or redundancy pruning without significantly affecting performance.

\noindent\textbf{Table synchronization.}
\short assumes the sender and receiver maintain consistent vector tables. In practice, synchronization can be maintained using simple mechanisms such as acknowledgments for fallback transmissions or periodic consistency checks. If divergence occurs, the system naturally recovers when a query misses and the full observation is retransmitted and reinserted into both tables.

\noindent\textbf{Embedding quality.}
The effectiveness of \short depends on the quality of the embedding space used for retrieval. Stronger encoders increase the probability that similar observations map to nearby vectors, improving hit rates and communication savings.

\noindent\textbf{Wide applicability.}
Although this work focuses on image and radar sensing, 
\tool is modality-agnostic. Any sensing modality that admits a meaningful embedding space, such as audio, 
LiDAR, or multivariate sensor streams, can potentially benefit from retrieval-based communication.

\section{Conclusion}
This paper introduces \tool{}, a communication paradigm that transmits sensor data by reference through vector retrieval over persistent memory. By exploiting the redundancy naturally present in sensing streams and leveraging abundant non-volatile storage on modern devices, \tool{} allows many observations to be communicated using compact references rather than full transmissions.

We design an efficient multi-level retrieval structure that enables large vector tables to reside in persistent memory while maintaining negligible lookup overhead on resource-constrained hardware. Through both theoretical modeling and a prototype implementation on an MCU platform, we show that modest local computation and storage access can substantially reduce communication cost, latency, and energy. Across 
three image datasets and one radar dataset, our evaluation demonstrates that \tool{} achieves a strong accuracy-bandwidth accuracy-energy tradeoff, outperforming representative compression-based and task-oriented communication baselines.

More broadly, this work suggests a shift in how sensing systems approach communication. Wireless spectrum is increasingly congested, and improving communication efficiency purely through physical layer advances is both difficult and the subject of decades of mature research. At the same time, sensing hardware continues to rapidly gain computational capability. This trend creates an opportunity to shift work from communication to local computation. By performing more processing on-device, systems can transmit significantly less data. This insight underlies our design. As sensing workloads continue to grow in scale and complexity, we believe \tool will become an increasingly important primitive for efficient edge–cloud systems.

\bibliographystyle{plain}
\bibliography{references}

\end{document}